\documentclass[aps,showpacs,prx,twocolumn,superscriptaddress, floatfix]{revtex4-2}

\usepackage[colorlinks,urlcolor=blue,linkcolor=blue,anchorcolor=blue,citecolor=blue]{hyperref}

\usepackage{amsmath}
\usepackage{graphicx}
\usepackage{subfigure}
\usepackage{color}
\usepackage{amsfonts}
\usepackage{tikz}
\usepackage{esint}
\usepackage{mathrsfs}
\usepackage{braket}
\usepackage{lettrine}
\usepackage[shortlabels]{enumitem}
\usepackage{lipsum}
\usepackage[ruled,vlined]{algorithm2e}

\DeclareMathOperator{\Tr}{Tr}

\begin{document}
\title{Evaluating many-body stabilizer R\'enyi entropy by sampling reduced Pauli strings: singularities, volume law, and nonlocal magic}

\author{Yi-Ming Ding}
\email{dingyiming@westlake.edu.cn}
\affiliation{State Key Laboratory of Surface Physics and Department of Physics, Fudan University, Shanghai 200438, China}
\affiliation{Department of Physics, School of Science and Research Center for Industries of the Future, Westlake University, Hangzhou 310030,  China}
\affiliation{Institute of Natural Sciences, Westlake Institute for Advanced Study, Hangzhou 310024, China}

\author{Zhe Wang}
\affiliation{Department of Physics, School of Science and Research Center for Industries of the Future, Westlake University, Hangzhou 310030,  China}
\affiliation{Institute of Natural Sciences, Westlake Institute for Advanced Study, Hangzhou 310024, China}

\author{Zheng Yan}
\email{zhengyan@westlake.edu.cn}
\affiliation{Department of Physics, School of Science and Research Center for Industries of the Future, Westlake University, Hangzhou 310030,  China}
\affiliation{Institute of Natural Sciences, Westlake Institute for Advanced Study, Hangzhou 310024, China}

\begin{abstract}
 We present a novel quantum Monte Carlo method for evaluating the $\alpha$-stabilizer R\'enyi entropy (SRE) for any integer $\alpha\ge 2$. 
By interpreting $\alpha$-SRE as partition function ratios, we eliminate the sign problem in the imaginary-time path integral by sampling \emph{reduced Pauli strings} within a \emph{reduced configuration space}, which enables efficient classical computations of $\alpha$-SRE and its derivatives to explore magic in previously inaccessible 2D/higher-dimensional systems.
We first isolate the free energy part in $2$-SRE, which is a trivial term.
Notably, at quantum critical points in 1D/2D transverse field Ising (TFI) models, we reveal nontrivial singularities associated with the \emph{characteristic function} contribution, directly tied to magic. 
Their interplay leads to complicated behaviors of $2$-SRE, avoiding extrema at critical points generally. 
In contrast, analyzing the volume-law correction to SRE reveals a discontinuity tied to criticalities, suggesting that it is more informative than the full-state magic. For conformal critical points, we claim it could reflect nonlocal magic residing in correlations.
Finally, we verify that $2$-SRE fails to characterize magic in mixed states (e.g. Gibbs states), yielding nonphysical results.
This work provides a powerful tool for exploring the roles of magic in large-scale many-body systems, and reveals intrinsic relation between magic and many-body physics.
\end{abstract}
\maketitle
% ::::::::::::::::::::::::::::::::::::::::::::::::::::::::::
% ::::::::::::::::::::::::::::::::::::::::::::::::::::::::::
% ::::::::::::::::::::::::::::::::::::::::::::::::::::::::::
% ::::::::::::::::::::::::::::::::::::::::::::::::::::::::::
\section{Introduction}

Characterizing many-body behaviors in quantum systems has been a central task since the last century~\cite{sachdev1999quantum, GirvinYang2019modern}. Despite the locality of interactions in most quantum matters, the intricate correlations that give rise to rich and exotic collective phenomena present formidable theoretical and computational hurdles for physicists. 
During the past decades, remarkable headways are made by introducing the theory of quantum information into many-body physics. Quantum entanglement, with its quantum nature and as probably the most representative quantum resource~\cite{eric2019quantumresource}, has turned out to relate closely to phase transitions, conformal field theories and topological orders~\cite{fazio2008entreview, zeng2019qinfo, nicolas2016ent}.

In recent years, another quantum resource called \emph{magic} or \emph{non-stabilizerness}, which is closely associated with the stabilizer formalism~\cite{arne2022magicresource,Veitch2014magicresource}, has also entered the field of vision in many-body physics~\cite{white2021cftmagic, winter2022mbm, poetri2023magic}.
According to the celebrated Gottesman-Knill theorem~\cite{gottesman1998, gottesman2004,Nielsen_Chuang_2010}, magic, rather than merely entanglement, is what necessarily enables quantum computation to outperform classical computation, as simulating a highly entangled stabilizer state requires only polynomial resources on a classical computer with Clifford protocols. 

From the aspect of computational complexity, magic captures a distinctive dimension of quantumness, which may reveal information of quantum states that goes beyond what entanglement alone can describe. 
Recent studies have shed light on the connections between magic and quantum criticality~\cite{winter2022mbm, white2021cftmagic,Sarkar_2020_1d_xy_rom, Haug2023stabilizerentropies, haug2023sremps,Haug2023stabilizerentropies, lami2023mpssampling_magic,poetri2023magic,poetri2024criticalbehaviorsof,Tarabunga2024generalizedRK}, quantum chaos~\cite{Leone2021quantumchaosis, leone2022sre}, and the AdS-CFT correspondence~\cite{white2021cftmagic,salvatore20221d_tfim_magic,kanato2022magic_rom_chaos}, yet much about the natures of magic remains unexplored. 
In addition, the investigations on many-body magic align with the perspective that ``entanglement is not enough" in exploring the intricate global properties of the inside geometry of black holes~\cite{Susskind2016ent_not_enough,Susskind2016complexity_blackhole,Stanford2014complexity_shock_wave,Roberts2015localizedShocks}.

To quantify magic in the many-body context, an ``appropriate" measure of magic is required. First, it should be monotone, which means it does not increase under Clifford protocols~\cite{arne2022magicresource,winter2022mbm,Haug2023stabilizerentropies}. Second, it should have good computability to enable analytical and numerical calculations. Consequently, measures such as the robustness of magic~\cite{mark2017robustness}, min-relative entropy of magic~\cite{Bravyi2019simulationofquantum, winter2022mbm}, and the extension of stabilizer norm~\cite{Tim2024monotone} are inappropriate due to the needs of complex minimization procedures.
% Consequently, \emph{mana}~\cite{Veitch2012mana, Veitch2014mana} and the \emph{$\alpha$-stabilizer R'enyi entropy (SRE)} ($\alpha\ge 2\in\mathbb{Z}$)~\cite{leone2022sre, Haug2023stabilizerentropies,leone2024sre} are often the choices. Although mana serves as a monotone for both the pure and mixed states, the computations of it generally require exponential resources.
In this paper, we consider the \emph{$\alpha$-stabilizer R\'enyi entropy ($\alpha$-SRE)} ($\alpha\ge 2\in\mathbb{Z}$), which is based on evaluating expectation values of Pauli strings~\cite{leone2022sre}.
The $\alpha$-SRE is a monotone for pure states, making it well-suited for exploring the magic of many-body ground states~\cite{Haug2023stabilizerentropies,leone2024sre}. 
In addition, it has close relation with the entanglement spectrum flatness for an arbitary bipartition~\cite{xhek2023flatness, emanuele2024flatness}. 

The contributions of this work span both algorithmic and physical levels. At the algorithmic level, we introduce the first unbiased Quantum Monte Carlo (QMC) algorithm capable of computing both the $\alpha$-SRE and its derivatives in large-scale and high-dimensional quantum many-body systems within polynomial time, as long as the Hamiltonian is sign-problem free. 
In fact, over the past years, several classical numerical tools have been developed to calculate the $\alpha$-SRE including the methods with matrix product states (MPS) and tree tensor networks~\cite{Haug2023stabilizerentropies, haug2023sremps, lami2023mpssampling_magic,poetri2023magic, poetri2024criticalbehaviorsof, poetri2024MPS}. However, these methods face limitations in two or higher dimensions and finite-temperature cases. A recent hybrid algorithm proposed in Ref.~\cite{zejun2024qmc_sre} addresses this issue by expressing the $\alpha$-SRE using the language of tensor network (TN) and then applying non-equilibrium QMC sampling~\cite{d2020entanglement}. However, their algorithm computes $\alpha$-SRE at only a single point per run and is incapable of extracting derivative information, yielding results with limited physical insight.

Technically, we prove that the QMC simulations of the $\alpha$-SRE, which relates to some generalized partition function, can be restricted in the \emph{reduced configuration space} by sampling \emph{reduced Pauli strings} (both terms are defined in Sec.~\ref{sec:algorithm}). Using this algorithm, it is straightforward to estimate the values and derivatives of $\alpha$-SRE by considering the partition function difference, in conjunction with Monte Carlo techniques including but not restricted to: the reweight-annealing method (importance sampling)~\cite{ding2024reweight,ding2024negativity,wz2024reweight,jiang2024high,wang2024addressing,wang2024probing,Ma2024defing,neal1993probabilistic,LPollet2008thmIntQMC}, thermodynamic integration~\cite{neal1993probabilistic, wukaixin2020negativity, FRENKEL2002167freeEnergyBook, XLMeng1998normalizing}, and the Bennett acceptance ratio method~\cite{Bennett1976interp, hahn2009acceptance}. 
The flexibility of our framework also allows for the integration of ideas from recent QMC algorithms for computing entanglement entropy, which could lead to further performance improvements~\cite{d2020entanglement,referee1_Isakov2011ent,referee2_melko2010mutual,referee3__stephan2012ent,referee4_Zhao2022ent,referee5_zhao2022dqcp,referee6_rong2008ent,referee7_emidio2015ent}.
The only limitation of our algorithm is that the Hamiltonian must be free of the sign problem, meaning that the weights in the partition function summation must be non-negative real numbers to allow a probabilistic interpretation—a common challenge for all QMC algorithms~\cite{Troyer2005sign,pan2024sign}. 
However, sign-problem-free models already host a rich variety of fundamental many-body phenomena, including spontaneous symmetry breaking, topological phase transitions, topological order, and conformal criticality. As such, our method plays a crucial role in advancing the understanding of quantum magic and complex quantum matter.

At the physical level and as a demonstration of our approach, we study the magic in the transverse field Ising (TFI) model with periodic boundary condition (PBC). The Hamiltonian is given by 
\begin{equation}\label{eq:tfim}
    H=-J\sum_{\langle ij\rangle}Z_iZ_j - h\sum_i X_i, 
\end{equation}
where $\langle ij\rangle$ denotes the nearest-neighbor site pairs, and $X$ and $Z$ refer to Pauli matrices. We investigate the ground-state magic of both 1D and 2D TFI model, as well as the finite-temperature $2$-SRE of the 2D TFI model. 

In previous studies, though the full-state magic of the ground state in general quantum models has shown not to be a reliable diagnostic for criticality~\cite{white2021cftmagic,Haug2023stabilizerentropies, haug2023sremps, lami2023mpssampling_magic,poetri2023magic, poetri2024criticalbehaviorsof, poetri2024MPS}, a further understanding of the relationship between the full-state magic and criticality remains open. 
In this work, we isolate the free energy part in $\alpha$-SRE for the first time, which contributes to a trivial singularity in its derivatives. 
In our study of ground states, we surprisingly find that the remaining nontrivial part—directly linked to the characteristic function (i.e., the distribution of squared expectation values of Pauli strings) and magic—also exhibits singularities in both the 1D and 2D TFI models. 
Consequently, the general behavior of the full-state magic quantified by $\alpha$-SRE is governed by the interplay between the contributions from the free energy and the characteristic function. The order of phase transition also plays a crucial role.
For examples, in the 1D TFI model, we observe that the ground-state magic peaks at the critical point, which is consistent with previous studies~\cite{Haug2023stabilizerentropies,poetri2023magic,poetri2024MPS}. In contrast, in the 2D TFI model, ground-state magic reaches its peak within the ferromagnetic phase. More broadly, this highlights a key reason why full-state magic does not always provide a clear signature of phases and criticalities in many-body systems.

Rather than focusing on the full-state magic, we suggest that its volume-law corrections would be more important as a nonzero volume-law correction is a necessary condition for the existence of \emph{nonlocal magic} residing in correlations. Such kind of magic is spread nonlocally and cannot be removed via local non-Clifford operations. 
Notably, considering the volume-law corrections does not require a computable mixed state monotone and has similar effect to the \emph{mutual magic} ~\cite{white2021cftmagic,Fliss2021long_range_magic,Sarkar_2020_1d_xy_rom,ning2022long_range_magic,frau2024magic_disentangling}, which we will detail in Sec.~\ref{sec:volume}. For both the 1D and 2D ground states of TFI models with finite system sizes, we observe discontinuous indications of the volume-law corrections at the critical point, which reflects the sudden transition of the ground-state magical structure across the phase transition point.

Additionally, we investigate the $2$-SRE of at the finite-temperature case in the 2D TFI model, observing that the singularity of the part of the characteristic function occurs outside the critical point, rendering it ineffective for characterizing the system. This accords with the understanding that $2$-SRE is not a well-defined measure for mixed-state magic~\cite{leone2022sre,Haug2023stabilizerentropies}. 
In fact, even for 1D systems, studies on the magic of Gibbs states for local Hamiltonians are scarce. Hopefully, extending our method to compute the mutual stabilizer R\'enyi entropy, discussed in Sec.~\ref{sec:summary}, could provide valuable new insights in the future.

The paper is organized as follows. In Sec.~\ref{sec:magic_intro}, we briefly review the magic resource theory including the definition of $\alpha$-SRE. Our core algorithm is introduced in Sec.~\ref{sec:algorithm}. With the algorithm, we can estimate the values and derivatives of $\alpha$-SRE in conjunction with the reweight-annealing and thermodynamic integration techniques, which we will discuss in Sec.~\ref{sec:pt_diff}. 
The numerical results for the magical behaviors related to many-body physics of the 1D and 2D TFI models are analyzed in Sec.~\ref{sec:results}, then in Sec.~\ref{sec:summary}, we provide a summary and further discussions on extending our algorithm to compute the $\alpha$-SRE of a reduced density matrix by tracing out the environment degrees of freedom.

% ::::::::::::::::::::::::::::::::::::::::::::::::::::::::::
% ::::::::::::::::::::::::::::::::::::::::::::::::::::::::::
\section{Quantum magic}\label{sec:magic_intro}
% ----------------------------------------
\subsection{Short review of stabilizer protocols} 
We focus our discussions on qubit systems in this paper, which can be extended to general qudit systems. We first define the Pauli group $\mathcal{G}_N$ as   
\begin{equation}
    \mathcal{G}_N := \bigg\{ \xi\otimes_j \sigma_j \bigg|  \sigma_j\in\{I,X,Y,Z\},\xi\in\{\pm 1,\pm i\}   \bigg\}, 
\end{equation}
where $X$, $ Y$, $Z$ are the three Pauli matrices and $I$ is the $2\times 2$ identity matrix. For an Abelian subgroup $\mathcal{S}\subset\mathcal{G}_N$ and some state space $V_{\mathcal{S}}$, if $\forall S\in\mathcal{S}$ and $\ket\psi\in V_{\mathcal{S}}$, $S\ket{\psi}=\ket{\psi}$ holds, then $\mathcal{S}$ is called a stabilizer of $V_{\mathcal{S}}$. 
In the Heisenberg picture, suppose $g$ is a generator of $\mathcal{S}$, and $U$ is some unitary operation acting on $\ket{\psi}$, then $U|\psi\rangle=Ug|\psi\rangle=UgU^{\dagger}U|\psi\rangle$, indicating that $UgU^{\dagger}$ stabilizes $U|\psi\rangle$. As the number of the generators $\{g_j\}$ is at most $\log |\mathcal{S}|$, then simulating the evolution of such kind of $\ket{\psi}$ only requires polynomial complexity in the Heisenberg picture.
To enable this, the evolved stabilizers must remain within the Pauli group under certain unitary operations. These unitaries are known as Clifford unitaries, which form the Clifford group $\mathcal{C}$. Each Clifford unitary can be efficiently generated from the Hadamard gate, the CNOT gate, and the Phase gate, all within polynomial complexity. Additionally, measurement operations in the computational basis can be incorporated into the circuit with polynomial complexity, formalized in the Gottesman-Knill theorem~\cite{gottesman1998, gottesman2004,Nielsen_Chuang_2010}. 
By introducing necessary operations, such as qubit discarding, we can construct a fully classical protocol referred to as the stabilizer protocol. The quantum states manipulated under this protocol hence have classical simulatability, and are called \emph{stabilizer states}.

To achieve universal quantum computation, it is necessary to introduce the quantum resources beyond the stabilizer protocol, which are known as the \emph{magic resources}~\cite{WangX2019magic,eric2019quantumresource,howard2017magic_resource,winter2022mbm,Zhan2022magicgate}. The greater the magic of a state, the further it deviates from classically simulable stabilizer states. As discussed earlier, a suitable measure of magic (a real map on density matrices) must satisfy the monotonicity condition. Suppose $\mathcal{M}$ is some measure, then for a given density matrix $\rho$ and any stabilizer protocol $\mathcal{E}$ that manipulates $\rho$, we must have $\mathcal M[\mathcal{E}(\rho)]\le \mathcal{M}(\rho)$. 

% ----------------------------------------
\subsection{$\alpha$-stabilizer R\'enyi entropy}\label{subsec:se}
For a pure state $\rho$, its $\alpha$-stabilizer R\'enyi entropy ($\alpha$-SRE) is defined as~\cite{leone2022sre}
\begin{equation}\label{eq:sre_pure}
    M_{\alpha}(\rho) := \frac{1}{1-\alpha} \log  \bigg[\frac{1}{2^N}\sum_{P\in\mathcal{P}_N}  
    {|\Tr(\rho P)|^{2\alpha}} \bigg], 
\end{equation}
where 
\begin{equation}\label{eq:pauligroup}
    \mathcal{P}_N := \bigg\{\otimes_{j=1}^N \sigma_j \bigg |  \sigma_j= I, X, Y, Z\bigg\}
\end{equation}
is a quotient group of the Pauli group $\mathcal{G}_N$, and the monotonicity is ensured for $\alpha\ge 2\in\mathbb{Z}$~\cite{Haug2023stabilizerentropies,leone2024sre}.
This definition can be viewed as the $\alpha$-R\'enyi entropy of a classical distribution 
$\Xi_P(\rho):=[\Tr(\rho P)]^2/2^N$, also called the \emph{characteristic function}~\cite{zhu2016clifford,leone2022sre}.
For convenience, we consider the case that $\alpha=2$ and also call $\mathcal{P}_N$ the Pauli group in this paper. 

The $2$-SRE $M_2(\rho)$ satisfies the following properties: (i) \emph{faithfulness}: ${M}_{\alpha}(\rho)=0$ iff $\rho$ has no magic; (ii) \emph{stability}: ${M}_{\alpha}(C\rho C^{\dagger})={M}_{\alpha}(\rho)$ for any Clifford unitary $C\in\mathcal{C}$ related to the Hilbert space; (iii) \emph{additivity}: ${M}_{\alpha}(\rho\otimes\rho')=M_{\alpha}(\rho) + M_{\alpha}(\rho')$ for arbitrary two pure states $\rho$ and $\rho'$.

The definition in Eq. (\ref{eq:sre_pure}) can be extended to mixed states as~\cite{leone2022sre}
\begin{equation}\label{eq:sre_mixed}
    \tilde{M}_{2}(\rho) := M_2(\rho)-S_2(\rho), 
\end{equation}
where 
\begin{equation}
    S_2(\rho):=-\log[\Tr(\rho^2)]
\end{equation}
is the entanglement 2-R\'enyi entropy of $\rho$. For pure states, $S_2(\rho)$ vanishes and $\tilde{M}_2(\rho)$ reduces to Eq.~(\ref{eq:sre_pure}). 
Note that $\tilde{M}_2(\rho)$ is not a well-defined measure for quantifying mixed-state magic, similar to entanglement entropy used in mixed states. 

% ----------------------------------------
\subsection{Generalized partition functions}\label{sec:gpt}
To compute the $2$-SRE, we first reformulate it in the language of QMC. 
Specifically, we employ the stochastic series expansion (SSE) method~\cite{sandvik1998stochastic, sandvik2003stochastic, Melko2013} in this paper.

Given a Hamiltonian $H$ and the inverse temperature $\beta$, the density matrix is $\rho=e^{-\beta H}/Z$, where $Z:=\Tr(e^{-\beta H})$ is the normalizing factor to guarantee $\Tr(\rho)=1$. %Please be careful that the $\rho$ simulated in QMC in the following is unnormalized. 
Similarly, we define $Z_2:=\Tr(e^{-2\beta H})$ such that $\Tr(\rho^2)=Z_2/Z^2$. Besides, we introduce 
\begin{equation}\label{eq:def_Q}
    Q := \sum_{P\in\mathcal{P}_N} |\Tr(e^{-\beta H} P)|^4 = \sum_{P\in\mathcal{P}_N} [\Tr(e^{-\beta H} P)]^4, 
\end{equation}
which can be viewed a generalized partition function. Then we have 
\begin{equation}\label{eq:pt_diff}
    \begin{split}
        \tilde{M}_2(\rho)=&-\log \bigg[\frac{1}{2^N}\frac{Q}{Z^4}\bigg] + \log\bigg[\frac{Z_2}{Z^2}\bigg]. 
    \end{split}
\end{equation}
Consequently, evaluating the $2$-SRE is equivalent to evaluating some partition function difference according to Eq.~(\ref{eq:pt_diff}).
The central challenge then lies in simulating $Q$ with QMC, which we will detail in the next section.

% ::::::::::::::::::::::::::::::::::::::::::::::::::::::::::
% ::::::::::::::::::::::::::::::::::::::::::::::::::::::::::
\section{Quantum Monte Carlo simulations of the generalized partition function}\label{sec:algorithm}
% ----------------------------------------
\subsection{Stochastic series expansion}
\begin{figure}[ht!] \centering
    \subfigure[] { \label{fig:1rep}
    \includegraphics[width=8.8cm]{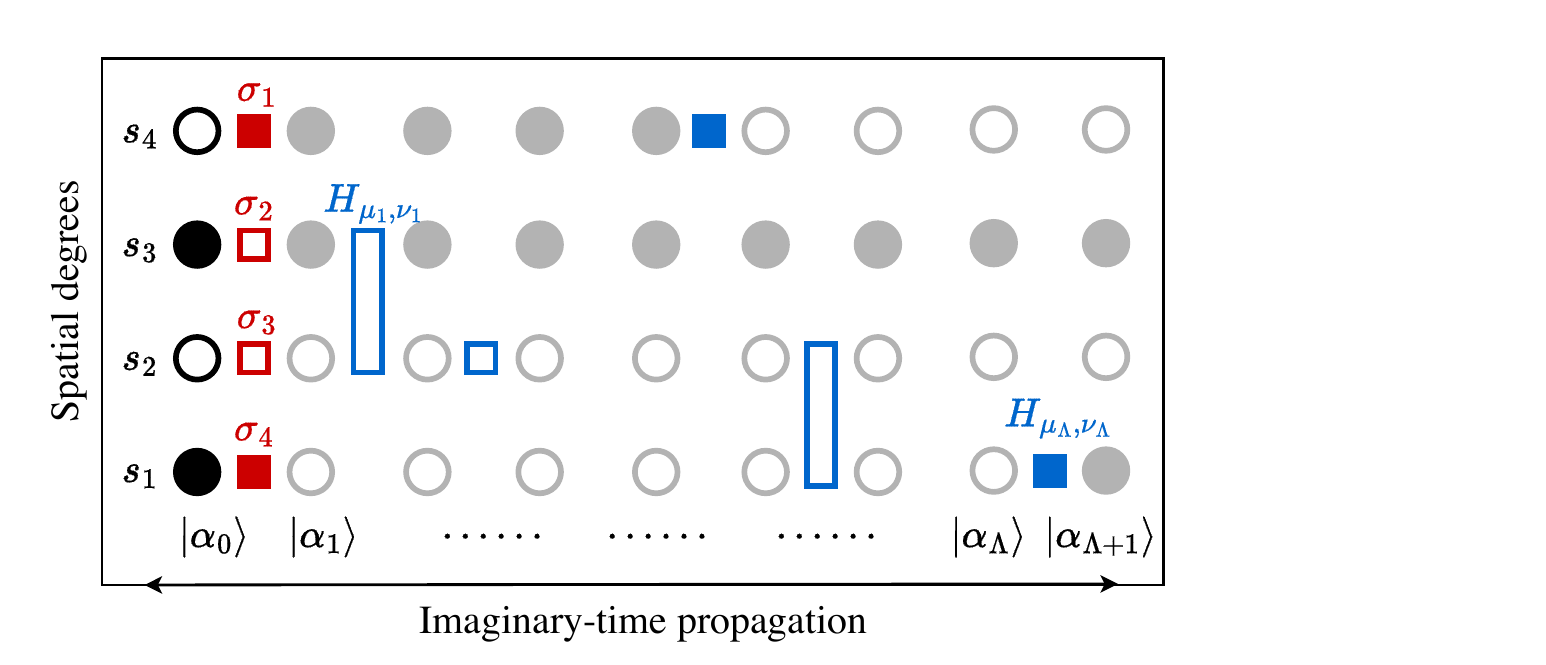}}
    \subfigure[] { \label{fig:4reps}
    \includegraphics[width=8.5cm]{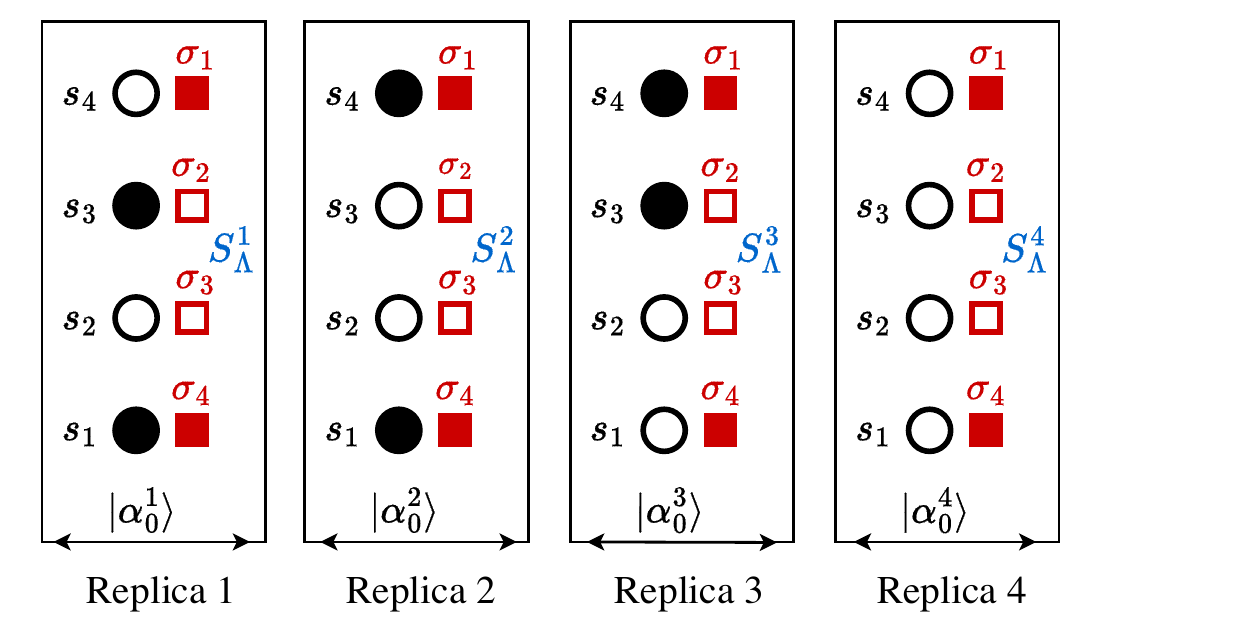}}
    \caption{ 
       (a) A configuration of $\Tr(e^{-\beta H} P)$ is shown in the language of SSE. The vertical axis represents the spatial degrees of freedom of the system, such as lattice sites $s_0$, $s_1$,$\cdots$. In this example, a qubit lies at each site, with empty and solid circles denoting $\ket{\uparrow}$ and $\ket{\downarrow}$, respectively. The horizontal axis indicates the $(\Lambda+1)$ {imaginary-time slices}. Associated to each slice, the state of qubits are represented by $\ket{\alpha_q}$. Then, $\ket{\alpha_0}$ is sequentially propagated by operators $P=\prod_j\sigma^j$, $H_{\mu_1,\nu_1}$, $H_{\mu_2,\nu_2}$, $\cdots$, and $H_{\mu_{\Lambda},\nu_{\Lambda} }$, evolving through the intermediate states $\ket{\alpha_{1}}$, $\ket{\alpha_2}$, $\cdots$, until it returns to $\ket{\alpha_{\Lambda+1}}\equiv \ket{\alpha_0}$. For diagonal (off-diagonal) $\sigma_j$ and $H_{\mu_q,\nu_q}$, we denote them with empty (solid) quadrangles in the diagram. 
       (b) A sampled configuration of $Q$ with four replicas of $\Tr(e^{-\beta H} P)$. The four replicas share the same $P$, while $\ket{\alpha_0^k}$ and $S_{\Lambda}^k$ can be different. Here we only show the time slice of $P$ for simplicity and the others are represented by $S_{\Lambda}^{k}$.
       }
        \label{fig}
\end{figure}

We start by considering $\Tr(e^{-\beta H} P)$ to facilitate understanding. The generalized partition function $Q$ just takes four replicas of $\Tr(e^{-\beta H} P)$ then sums over all possible Pauli string $\{P\}$, which will be discussed later. Following the standard procedure of SSE~\cite{sandvik1998stochastic, sandvik2003stochastic, Melko2013,yan2019sweeping,yan2022improved}, we first apply the Taylor expansion to get 
\begin{equation} 
    \begin{split}
        \Tr(e^{-\beta H} P) =& \sum_n\frac{\beta^n}{n!} \Tr[(-H)^nP] . 
    \end{split}
\end{equation}
Then we rewrite $H=-\sum_{\mu,\nu}H_{\mu,\nu}$, where each $H_{\mu,\nu}$ is a local operator with $\mu$ to denote its type (diagonal or off-diagonal under a given representation) and $\nu$ specifying the spatial degree (bond or site) on which the operator acts on. Then  
\begin{equation}\label{eq:series}
    \Tr(e^{-\beta H} P) =\sum_n\sum_{S_n}\frac{\beta^n}{n!}
    \Tr\bigg\{
    \bigg[\prod_{q=1}^{n}H_{\mu_p,\nu_p}\bigg] P     
    \bigg\},
\end{equation}
where each $S_n$ specifies an operator sequence or string $\prod_{q=1}^{n}H_{\mu_q,\nu_q}$ with length $n$. 
For practical simulations, series (\ref{eq:series}) must be truncated at some sufficiently large power $\Lambda$. After filling in $(\Lambda-n)$ null operators (also denoted by $H_{\mu,\nu}$ with some special $\mu$), we have 
\begin{equation}\label{eq:expansion}
    \Tr(e^{-\beta H} P) =\sum_{\alpha}\sum_{S_{\Lambda}} W(P;{S}_{\Lambda},\ket{\alpha_{0}}),
\end{equation}
where
\begin{equation}\label{eq:prop}
    \begin{split}
        & W(P;{S}_{\Lambda},\ket{\alpha_0}) \\&=\frac{\beta^n(\Lambda-n)!}{\Lambda!}
\bigg[\prod_{q=1}^{\Lambda}\langle \alpha_{q+1}|H_{\mu_q, \nu_q}|\alpha_{q}\rangle\bigg] \bra{\alpha_1}P\ket{\alpha_0}
    \end{split}
\end{equation}
are sampling weights, and the summation on $\alpha$ includes totally $(\Lambda+1)$ sets of basis $\ket{\alpha_q}$ propagated by $P$ and $\{H_{\mu_q,\nu_q}\}$. In addition, there is a PBC ($\ket{\alpha_0}\equiv \ket{\alpha_{M+1}}$) along the imaginary-time axis because of the trace operation.
Fig.~\ref{fig:1rep} presents a schematic diagram of a configuration associated to $W(P;S_{\Lambda},\ket{\alpha_0})$, which apparently can be uniquely determined by $P$, $S_{\Lambda}$ and $\ket{\alpha_0}$.

Similarly, we can apply SSE to $Q$ to obtain 
\begin{equation}\label{eq:Q_weight}
Q=\sum_{P\in\mathcal{P}_N}\sum_{\alpha^k}\sum_{{S}^k_{\Lambda}}\bigg[\prod_{k=1}^4W(P;{S}_{\Lambda}^k,\ket{\alpha_0^k})\bigg],
\end{equation}  
where
\begin{equation}\label{eq:prop_rep}
    \begin{split}
        & W(P;{S}_{\Lambda}^k,\ket{\alpha^k_0}) \\&=\frac{\beta^n(\Lambda-n)!}{\Lambda!}
\bigg[\prod_{q=1}^{\Lambda}\langle \alpha^k_{q+1}|H^k_{\mu_q, \nu_q}|\alpha^k_{q}\rangle\bigg] \bra{\alpha^k_1}P\ket{\alpha^k_0}
    \end{split}
\end{equation}
and the superscript $k$ signifies the replica number. 
Notice that $S_{\Lambda}^k$ or $\ket{\alpha_0^k}$ can differ across replicas, thus each configuration of $Q$ is uniquely determined by $P$, $\{S_{\Lambda}^k\}$ and $\{\ket{\alpha_0^k}\}$ (see Fig.~\ref{fig:4reps}). 

For $H$ representing some qubit system, $\alpha_q^k$ are typically chosen in the basis of Pauli operator $Z$. 
Since $Z=\text{diag}(1,-1)$ introduces a minus sign after propagating one $\ket{\downarrow}$ in $\ket{\alpha_0}$ to another $\ket{\downarrow}$ in $\ket{\alpha_1}$, we will encounter negative weights in $\prod_{k}W(P;{S}_{\Lambda+1}^k,\alpha_0^k)$, which results in the notorious sign problem~\cite{Sugar1990exp,takasu1986monte,hatano1992representation,PhysRevB.92.045110,Zhou2019Universal}. Similar bother also happens on the Pauli operator $Y$.
Fortunately, this issue can be avoided by leveraging the symmetry of Pauli strings by performing simulations in a reduced configuration space, as introduced next.

% ----------------------------------------
\subsection{Reduced configuration space}
First, we replace $Y$ in the Pauli group (\ref{eq:pauligroup}) with 
\begin{equation} 
    \tilde{Y}=\begin{bmatrix}
        0 & -1 \\ 1 & 0
    \end{bmatrix}
\end{equation} 
to remove imaginary numbers. This makes no difference on simulating $Q$ according to the definition (\ref{eq:sre_pure}).

For convenience, let us denote by $\mathfrak{C}$ the parts of each configuration of $Q$ expressed in Eq.~(\ref{eq:Q_weight}) determined by $\{{S}_{\Lambda}^k\}$ and $\{\ket{\alpha_0^k}\}$. To identify all possible configurations, we can start by selecting a $\mathfrak{C}$, then traverse all associated $P$ before moving to the next $\mathfrak{C}$.
Notably, to ensure the PBC in the imaginary-time axis, the type of each $\sigma_j$ (diagonal or off-diagonal) in $P$ is also constrained by $\mathfrak{C}$. 

A simplified problem is to consider a single replica and a single qubit, i.e. we consider the partition function $\sum_{P\in\mathcal{P}_1}\Tr(e^{-\beta H} P)$. If $P$ is off-diagonal under a given $\mathfrak{C}$ with $\ket{\alpha_0}=\ket{\uparrow}$, then $X$ and $\tilde{Y}$ both contribute a positive sign after propagating $\ket{\alpha_0}$, with their corresponding configurations sharing the same weight. On another hand, if $\ket{\alpha_0}=\ket{\downarrow}$, $\tilde{Y}$ yields a negative sign, in which case, the two weights are equal in magnitude but opposite in sign. As a result, in the summation like Eq.~(\ref{eq:Q_weight}), these two terms cancel each other, and this $\mathfrak{C}$ does not appear in the valid configuration space. This can be easily generalized to an arbitrary type of $P$ and $N$ qubits, 
%and an example of two qubits is shown in Table \ref{table:1rep2spins}.
and effectively, $\mathcal{P}_N=\{\prod_{j}\sigma_j|\sigma_j\in\{(X+\tilde{Y}),(I+Z)\}\}$ in this case. 

With a single replica, any valid $\mathfrak{C}$ must satisfy $\ket{\alpha_0}=\ket{\uparrow\cdots\uparrow}$. We refer to the corresponding valid configuration space as the \emph{reduced configuration space}, in contrast to the full space that includes configurations with negative weights. 
Furthermore, it is unnecessary to determine the specific Pauli matrix for each $\sigma_j$; we only need to distinguish whether it is diagonal. 
This leads to the concept of \emph{reduced Pauli string}, such as $(X+\tilde{Y})$ for $\mathcal{P}_1$, which has considered the combined effects of multiple Pauli strings. This confirms we are on the right track. Using a similar idea, we can now proceed to tackle our target partition function $Q$. 

% \begin{table}[htbp]
% 	\centering 
% 	\begin{tabular}{|c|c|c|c|c|c|}  
% 		\hline  
% 		 & $X_1\otimes X_2$ & $X_1\otimes \tilde{Y}_2$ & $\tilde Y_1\otimes X_2$ & $\tilde Y_1\otimes \tilde Y_2$ & Sum\\ 
% 		\hline
% 		$\ket{\uparrow\uparrow}$ & $+$ & $+$ & $+$ & $+$ &  $+$\\
%         \hline
% 		$\ket{\uparrow\downarrow}$ & $+$ & $-$ & $+$ & $-$ & $0$\\
%         \hline
% 		$\ket{\downarrow\uparrow}$ & $+$ & $+$ & $-$ & $-$ & $0$\\
% 		\hline
%         $\ket{\downarrow\downarrow}$ & $+$ & $-$ & $-$ & $+$ & $0$\\
% 		\hline
% 	\end{tabular}
%     \caption{sadas}
%     \label{table:1rep2spins}
% \end{table}

For simplicity, we still start by considering one qubit and assume $P$ to be off-diagonal. It is easy to verify that as long as an even number of $\ket{\alpha_0^k}$ is in $\ket{\uparrow}$, the corresponding $\mathfrak{C}$ will be valid, due to the multiplication of $W(P;S_{\Lambda}^k,\ket{\alpha_0^k})$ in Eq.~(\ref{eq:Q_weight}). Otherwise, the symmetric contributions from $X$ and $\tilde{Y}$ would analogously make $\mathfrak{C}$ absent in the reduced configuration space.  

Concretely, for the $N$-qubit case, we denote $\ket{\alpha_0^k}=\ket{q^k_1,\cdots,q^k_N}$, where $\ket{q^k_{j}}\in\{\ket{\uparrow},\ket{\downarrow}\}$ is the state of the $j$th qubit in replica $k$, then for any site $j$, there must be an even number of $\ket{q_j^k}$ in $\ket{\uparrow}$ state to ensure a valid $\mathfrak{C}$. The specific forms of these reduced Pauli strings in this case are also unimportant, and only their types and effects on $\ket{\alpha_0^k}$ matter, flipping or leaving the qubits unchanged. 

We now rewrite Eq.~(\ref{eq:Q_weight}) as 
\begin{equation}\label{eq:Q_weight2}
Q=\sum_{\tilde{P}\in\tilde{\mathcal{P}}_N}\sum_{\tilde{\alpha}^k}\sum_{\tilde{S}^k_{\Lambda}}\bigg[\prod_{k=1}^4W(\tilde P;\tilde{S}_{\Lambda}^k,\ket{\tilde\alpha_0^k})\bigg],
    \end{equation}  
where the tildes indicate that we are considering the reduced configuration space and reduced Pauli strings.

% ----------------------------------------
\subsection{The algorithm}\label{subsec:algorithm}
Compared to a standard SSE algorithm which involves updates on operators $\{H^k_{\mu_q,\nu_q}\}$ and spins $\{\ket{\alpha_0^k}\}$~\cite{sandvik1998stochastic, sandvik2003stochastic, OFS2002directed,Melko2013,sandvik2019stochastic,sandviki2011tutorial}, two modifications are made in our algorithm: 
(i) the updates on $\ket{\alpha_0^k}$ must ensure that even number of $\ket{q_j^k}$ are in $\ket{\uparrow}$; 
(ii) the reduced Pauli string $\tilde{P}=\prod_j\tilde{\sigma}_j$, where $\tilde{\sigma}_j$ is diagonal or off-diagonal $\sigma_j$, must also be updated. 
We summarize the key components of our algorithm below, which is broadly applicable to any sign-problem-free models in SSE simulations.

% ................................................
\subsubsection{Local update} 
The local update for $Q$ follows a similar approach to that in a conventional SSE algorithm, where $\tilde{P}$, $\{\tilde{\alpha}_0^k\}$, and the off-diagonal operators remain unchanged. For each time slice in each replica, consider updating the null operator (corresponding to an empty time slice) to a diagonal one and vice versa, following the Metropolis-Hasting algorithm~\cite{Metropolis1953,Hasting1970}.

% ................................................
\subsubsection{Nonlocal update on the reduced Pauli string and operators} 
In SSE simulations, nonlocal updates, such as (directed) loop or branching cluster updates, are used to reduce autocorrelations during sampling. These updates alternate between diagonal and off-diagonal operators, which can improve the efficiency of simulations. 

A simple way to update $\tilde{P}=\prod_j\tilde{\sigma}_j$ is by treating  $\{\tilde{\sigma}_j\}$ as effective single-site diagonal and off-diagonal operators.
Then, each $\sigma_j$ can be involved in some nonlocal collection (a loop or cluster), and be updated together with those operators. 
Through this way, once a collection across the imaginary-time boundary is updated, the corresponding $\ket{q^k_j}$ should also be flipped (updated). 
On another hand, for each site $j$, we must always have even number of $\ket{q_j^k}$ in $\ket{\uparrow}$, then such an update of the collection should flip even number of $\ket{q_j^k}$ simultaneously. This suggests that the collections for nonlocal updates in this case be formed across different replicas. For a concrete example, readers may refer to Appendix~\ref{appx:tfim}, where we detail the nonlocal update scheme for the TFI model and analyze its autocorrelation times.

% \begin{algorithm}\label{alg:local}
%     \caption{Local update} 
%     \For{$k=1,2,3,4$}
%     {   
%         $\ket{\alpha'}\gets \ket{\tilde{\alpha}^k_0}$ \; 
%         $\ket{\alpha'}\gets P\ket{\alpha'}$\; 
%         \For{$q=1,\cdots,M$}{
%             \If{$H^k_{\mu_q,\nu_q}$ is off-diagonal}{
%                 $\ket{\alpha'}\gets H^k_{\mu_q,\nu_q}\ket{\alpha'}$
%             }
%             \Else{
%                 $\ket{\alpha'}$ unchanges\;
%                 \If{$H^k_{\mu_q,\nu_q}$ is an identity operator}{
%                     Consider changing it to a diagonal one with Metropolis algorithm\; 
%                 }  
%                 \Else{
%                     Consider changing it to an identity one with Metropolis algorithm\; 
%                 } 
%             }
%         }
%     } 
% \end{algorithm}

% \begin{algorithm}
%     \caption{SSE simulations for $Q$}
%     \KwIn{$n \geq 0$}
%     \KwOut{$y = x^n$}
%     Diagonal update 
%     $X \gets x$\;
%     $N \gets n$\;
%     \While{$N \neq 0$}{
%       \If{$N$ is even}{
%         $X \gets X \times X$\;
%         $N \gets \frac{N}{2}$\;
%       }
%       \ElseIf{$N$ is odd}{
%         $y \gets y \times X$\;
%         $N \gets N - 1$\;
%       }
%     }
%   \end{algorithm}

% ::::::::::::::::::::::::::::::::::::::::::::::::::::::::::
% ::::::::::::::::::::::::::::::::::::::::::::::::::::::::::
% ::::::::::::::::::::::::::::::::::::::::::::::::::::::::::
% ::::::::::::::::::::::::::::::::::::::::::::::::::::::::::
\section{Partition function difference}\label{sec:pt_diff}
% ----------------------------------------
\subsection{Reweight-annealing and importance sampling}\label{eq:re_an_intro} 
As our primary interest lies in how $\tilde{M}_2$ and its derivatives change with some physical parameter $\lambda$, we consider 
\begin{equation}\label{eq:pt_diff2}
    \tilde{M}_2(\lambda)=-\log Q(\lambda) + 2\log Z(\lambda) + \log Z_2(\lambda) + N\log 2
\end{equation}
and use the reweight-annealing (ReAn) method based on importance sampling~\cite{ding2024reweight,ding2024negativity,wz2024reweight,jiang2024high,wang2024addressing,wang2024probing,Ma2024defing,neal1993probabilistic,LPollet2008thmIntQMC}. 

First we select a reference point $\lambda_0$ for which the value of $\tilde M_2(\lambda_0)$ is already known. For example, in the TFI model (\ref{eq:tfim}), the limit $\lambda_0\equiv J=0$ corresponds to a product state with no magic. Then by considering the difference $[\tilde{M}_2(\lambda)-\tilde{M}_2(\lambda_0)]$ (assume $\lambda_0<\lambda$ without loss of generality), we have 
\begin{equation}\label{eq:ti_start}
    \begin{split}
        &\tilde{M}_2(\lambda) =\log\frac{Q(\lambda_0)}{Q(\lambda)} - 2\log\frac{Z(\lambda_0)}{Z(\lambda)}-\log \frac{Z_2(\lambda_0)}{Z_2(\lambda)}.
    \end{split}
\end{equation}
Since the distribution overlap between $Q(\lambda_0)$ and $Q(\lambda)$ (similar for $Z$ and $Z_2$) can be significantly small, practically, to enhance the simulation efficiency, interpolations are needed. We consider  
\begin{equation}\label{eq:interpolations}
    \begin{split}
        &\frac{Q(\lambda_0)}{Q(\lambda)}=\prod_{k=1}^m\frac{Q(\lambda_{k-1})}{Q(\lambda_{k})} ,\\ 
    &\frac{Z(\lambda_0)}{Z(\lambda)}=\prod_{k=1}^m\frac{Z(\lambda_{k-1})}{Z(\lambda_{k})} ,\\ 
    &\frac{Z_2(\lambda_0)}{Z_2(\lambda)}=\prod_{k=1}^m\frac{Z_2(\lambda_{k-1})}{Z_2(\lambda_{k})},
    \end{split}
\end{equation}
by dividing $[\lambda_0,\lambda]$ into $m$ subintervals, where $\lambda_m\equiv \lambda$.
Apparently the efficiency of Eq.~(\ref{eq:interpolations}) depends on (i) the number of $m$ required; (ii) the relative error for estimating $[Q(\lambda_{k-1})/Q(\lambda_k)]$. Neither of these two things should scale exponentially with system size in order to enable efficient simulations. 
To satisfy the two requirements, we use the annealing scheme introduced in Ref.~\cite{ding2024reweight}. 
In brief, it sets $[Q(\lambda_{k-1})/Q(\lambda_k)]\equiv \epsilon$, where $\epsilon<1$ is a constant for any parameter point and system size. Then, it can be proved that the required number of interpolations scales $m\sim (\lambda-\lambda_0)L^d/|\log \epsilon|$, which is polynomial, where $L$ is the length of the system and $d$ is the space dimension.
The relative error for estimating each $[Q(\lambda_{k-1})/Q(\lambda_k)]$ in this scheme then is bounded by $\sqrt{(\epsilon^{-1} - 1)/{N_{\text{MC}}}}$, with $N_{\text{MC}}$ to denote the number of Monte Carlo steps, which does not diverge exponentially.

If $\lambda\equiv \beta$, which is the inverse temperature, then the estimators are 
\begin{equation}\label{eq:estimator:beta}
    \begin{split}
        &\frac{Q(\beta_{k-1})}{Q(\beta_k)} =\bigg\langle \bigg(\frac{\beta_{k-1}}{\beta_k} \bigg)^{n_{\text{tot}}}\bigg\rangle_{\beta_k, Q}, \\ 
        &\frac{Z(\beta_{k-1})}{Z(\beta_k)} =\bigg\langle \bigg(\frac{\beta_{k-1}}{\beta_k} \bigg)^{n_{\text{tot}}}\bigg\rangle_{\beta_k, Z} ,\\ 
        &\frac{Z_2(\beta_{k-1})}{Z_2(\beta_k)} =\bigg\langle \bigg(\frac{\beta_{k-1}}{\beta_k} \bigg)^{n_{\text{tot}}}\bigg\rangle_{\beta_k, Z_2} ,
    \end{split}
\end{equation}
where the subscripts denote the simulation is at $\beta_k$ for simulating the corresponding partition function and $n_{\text{tot}}$ denotes the total number of operators (excluding the null operators) in the SSE configuration. The reference point is taken to be $\beta_0=0$, where the value of $2$-SRE is zero.

Similarly, if we take $\lambda\equiv J$ in the Hamiltonian of TFI model (\ref{eq:tfim}), we have 
\begin{equation}\label{eq:estimator:j}
    \begin{split}
        &\frac{Q(J_{k-1})}{Q(J_k)} =\bigg\langle \bigg(\frac{J_{k-1}}{J_k} \bigg)^{n_{J,\text{tot}}}\bigg\rangle_{J_k, Q}, \\ 
        &\frac{Z(J_{k-1})}{Z(J_k)} =\bigg\langle \bigg(\frac{J_{k-1}}{J_k} \bigg)^{n_{J,\text{tot}}}\bigg\rangle_{J_k, Z}, \\ 
        &\frac{Z_2(J_{k-1})}{Z_2(J_k)} =\bigg\langle \bigg(\frac{J_{k-1}}{J_k} \bigg)^{n_{J,\text{tot}}}\bigg\rangle_{J_k, Z_2} , 
    \end{split}
\end{equation}
where $n_{J,\text{tot}}$ denotes the total number of operators related to interacting terms $J$ in TFI model.

The derivations for the relations (\ref{eq:estimator:beta}) and (\ref{eq:estimator:j}) can be found in Appendix~\ref{appx:estimator}.

% ----------------------------------------
\subsection{Thermodynamic integration} 
Another popular way to estimate the partition function difference is the thermodynamic integration (TI)~\cite{neal1993probabilistic, wukaixin2020negativity, FRENKEL2002167freeEnergyBook, XLMeng1998normalizing}. In general, TI considers two systems $A$ and $B$, which can differ in their manifolds and parameter, with effective potential energies $E_A$ and $E_B$, respectively. 
To evolve $A$ to $B$, an extended system is introduced with an effective potential energy $E(\lambda)$, where $\lambda\in[\lambda_A,\lambda_B]$. The function $E(\lambda)$ is constructed such that $E(\lambda_A)=E_A$ and $E(\lambda_B)=E_B$. 
For example, suppose $A$ and $B$ share the same inverse temperature $\beta$, then for the extended system, $Z(\lambda)=\sum_l e^{-\beta E_l(\lambda)}$, and 
\begin{equation}\label{eq:ti_diff}
    \begin{split}
        \frac{d\log Z(\lambda)}{d\lambda} 
        %=& -\frac{\beta}{Z(\lambda)}\sum_c e^{-E_c(\lambda)/T}\frac{d E_c(\lambda)}{d \lambda} \\ 
        =&-\beta\bigg\langle \frac{dE(\lambda)}{d\lambda}  \bigg\rangle_{\lambda}.
    \end{split}
\end{equation}
Then integrating Eq.~(\ref{eq:ti_diff}), we achieve 
\begin{equation}\label{eq:ti2}
    \log \frac{Z_B}{Z_A}=\int_{\lambda_A}^{\lambda_B} -\beta \bigg\langle {d E(\lambda)}\bigg\rangle_{\lambda}.
\end{equation}
Note that Eq.~(\ref{eq:ti2}) also indicates how to calculate the derivatives of $\log(Z_B/Z_A)$ with parameter $\lambda$.

% Before we apply TI to calculating the $2$-SRE, we provide some comments to it: (i) the integration path specified by $\lambda$ can be unphysical, such as that in the snake algorithm and the nonequilibrium algorithms for calculating the entanglement entropies~\cite{Forcrand2001snakeAlgorithm, MCaselle2003thmIntExp, Jonathan2020ent, Zhao2022ent}. As a result, a good design of the unphysical path can improve the efficiency on evaluating $\log (Z_B/Z_A)$; (ii) since we need a bulk of interpolations $\lambda_j\in[\lambda_A,\lambda_B]$ to reduce the error of numerical integral, if the intermidiate points are physical or meaningful, the integration can actually provide the information of the full path of $\lambda$, equivalent to that in the ReAn method in the last subsection; (iii) Eq.~(\ref{eq:ti2}) also indicates how to calculate the derivatives of $\log(Z_B/Z_A)$ with parameter $\lambda$; (iv) the integration path can be nonequilibrium or non-adiabatic, then by applying the Jarzynski equality~\cite{jarzynski1997}, we can still estimate the value of $\log (Z_B/Z_A)$.

If we restrict the intermediate process of the integration to be adiabatic, repeated thermalization procedures can be saved by taking the final state from the last simulation as the initial state for the next simulation. This is a trick of simulated annealing~\cite{Kirkpatrick1983SA} and also applies to the ReAn method. In this work, we use TI to calculate the derivatives of $2$-SRE by identifying $Z_B\equiv Q(\lambda_1)$ and $Z_A\equiv Q(\lambda_2)$ (similarly for $Z$ and $Z_2$). We use ReAn to compute the values of $\tilde{M}_2$ instead of TI because empirically TI requires more interpolations than ReAn, as the former one has to consider numerical integration errors.

According to Eq.~(\ref{eq:ti_start}), the first-order derivatives of the $2$-SRE with parameter $J$ and $\beta$ can be easily achieved as 
\begin{align}
    \frac{d\tilde{M}_2(J)}{dJ} = \langle E\rangle_{J,Q}-2\langle E\rangle_{J,Z} -\langle E\rangle_{J,Z_2} \label{eq:first_order_J} , \\  
    \frac{d\tilde{M}_2(\beta)}{d\beta}=
    \langle E\rangle_{\beta,Q}-2\langle E\rangle_{\beta,Z} -\langle E\rangle_{\beta,Z_2},
    \label{eq:first_order_beta}
\end{align}
respectively, where the effective potential energies are 
\begin{equation}
    \begin{split}
        &\langle E\rangle_{J',Q} := -
        \frac{\langle n_{J,\text{tot}} \rangle_{J',Q}}{J'}, \\ 
        &\langle E\rangle_{J',Z} := -
        \frac{\langle n_{J,\text{tot}} \rangle_{J',Z}}{J'}, \\ 
        &\langle E\rangle_{J',Z_2} := -
        \frac{\langle n_{J,\text{tot}} \rangle_{J'Z_2}}{J'}, \\ 
    \end{split}
\end{equation}
and 
\begin{equation}
    \begin{split}
        &\langle E\rangle_{\beta,Q} := -
        \frac{\langle n_{\text{tot}} \rangle_{\beta,Q}}{\beta}, \\ 
        &\langle E\rangle_{\beta,Z} := -
        \frac{\langle n_{\text{tot}} \rangle_{\beta,Z}}{\beta}, \\ 
        &\langle E\rangle_{\beta,Z_2} := -
        \frac{\langle n_{\text{tot}} \rangle_{\beta,Z_2}}{\beta}, \\ 
    \end{split}
\end{equation}

The TI relations for higher-order derivatives of $\tilde{M_2}$ can also be estimated. For example, the second-order derivatives are 
\begin{align}
    &\frac{d^2\tilde{M}_2(J)}{dJ^2}=-\frac{1}{J^2}\bigg[
        C_{J,Q} - 2C_{J, Z} - C_{J,Z_2}    
    \bigg]\label{eq:second_order_J}, \\
    &\frac{d^2\tilde{M}_2(\beta)}{d\beta^2}=-\frac{1}{\beta^2}\bigg[
        C_{\beta,Q} - 2C_{\beta, Z} - C_{\beta,Z_2}   
    \bigg],
    \label{eq:second_order_beta}
\end{align}
where 
\begin{equation}
    \begin{split}
        &\langle C\rangle_{Q,J} := \langle n_{J,\text{tot}}(n_{J,\text{tot}}-1)\rangle_{J,Q} - \langle n_{J,\text{tot}}\rangle^2_{J, Q} , \\ 
        &\langle C\rangle_{Z,J} := \langle n_{J,\text{tot}}(n_{J,\text{tot}}-1)\rangle_{J,Z} - \langle n_{J,\text{tot}}\rangle^2_{J, Z} , \\ 
        &\langle C\rangle_{Z_2,J} := \langle n_{J,\text{tot}}(n_{J,\text{tot}}-1)\rangle_{J,Z_2} - \langle n_{J, \text{tot}}\rangle^2_{J, Z_2},
    \end{split}
\end{equation}
and 
\begin{equation}
    \begin{split}
        &\langle C\rangle_{Q,\beta} := \langle n_{\text{tot}}(n_{\text{tot}}-1)\rangle_{\beta,Q} - \langle n_{\text{tot}}\rangle^2_{\beta, Q} ,\\ 
        &\langle C\rangle_{Z,\beta} := \langle n_{\text{tot}}(n_{\text{tot}}-1)\rangle_{\beta,Z} - \langle n_{\text{tot}}\rangle^2_{\beta, Z} ,\\ 
        &\langle C\rangle_{Z_2,\beta} := \langle n_{\text{tot}}(n_{\text{tot}}-1)\rangle_{\beta,Z_2} - \langle n_{\text{tot}}\rangle^2_{\beta, Z_2} .\\ 
    \end{split}
\end{equation}

% --------------------------------------------
\section{Numerical results}\label{sec:results}
% =======================================
\subsection{Ground-state $2$-SRE and the derivatives}
\begin{figure}[ht!] \centering
    \includegraphics[width=8cm]{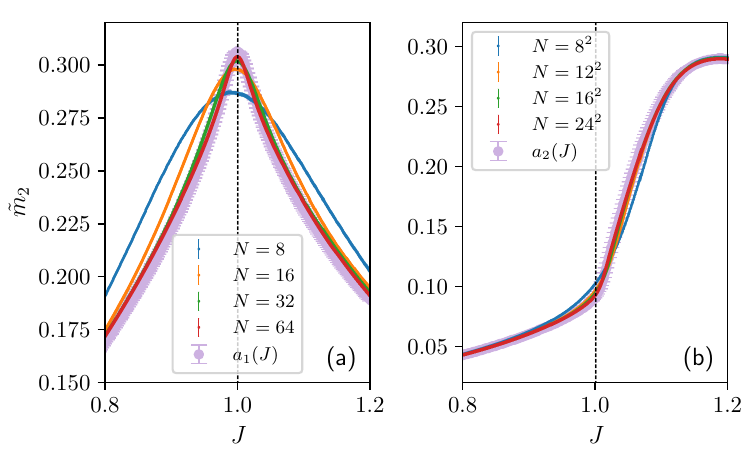}
    \caption{$\tilde{m}_2$ as a function of the coupling strength $J$ of the TFI model: (a) for the 1D ring with length $L$, fixing $h=1$ and $\beta=L$ to extrapolate to the ground state; (b) for the 2D $L\times L$ square lattice with PBC, fixing $h=3.04438$ and $\beta=L$ to extrapolate to the ground state. In addition, $a_d$ ($d=1,2$) stand for the volume-law coefficients in Eq.~(\ref{eq:fittttt}).
    }
    \label{fig:gs_sre}
\end{figure}
As many-body magic is typically an extensive quantity that follows the volume law~\cite{winter2022mbm}, its density is a commonly studied quantity~\cite{white2021cftmagic,poetri2023magic,poetri2024MPS,poetri2024criticalbehaviorsof}. For the $2$-SRE, we define the magic ($2$-SRE) density to be $\tilde{m}_2 := \tilde{M}_2 / N$. 

For the 1D TFI model, we consider a ring of length $L$ with the parameter $h$ fixed at $1$ in Eq.~(\ref{eq:tfim}), such that the quantum critical point is located at $J_c=1$. For the 2D TFI model on a $L\times L$ square lattice, we similarly fix $h=3.04438$ to let $J_c\approx 1$~\cite{dyj2022tfim}.

As shown in Fig.~\ref{fig:gs_sre}(a), the magic density $\tilde{m}_2$ of 1D TFI model reaches its maximum at the critical point, which aligns with previous results obtained via MPS and TN methods~\cite{Haug2023stabilizerentropies,poetri2023magic,poetri2024MPS}. 
In contrast, the magic of 2D TFI model increases  monotonically across the critical point, thereafter the magic for the 2D TFI model maximizes within the ferromagnetic (FM) phase and then decays to zero since the limit $J\to\infty$ corresponds to a stabilizer state. %\zyan{Local phase of spins have been enhaced near the critical point?}

This accords with the results reported by sampling generalized Rokhsar-Kivelson wavefunctions of the 2D quantum ferromagnet with the same phase transition in Ref.~\cite{Tarabunga2024generalizedRK}. However, their findings for the 1D quantum ferromagnet~\cite{Tarabunga2024generalizedRK} differ significantly from our 1D TFI model results. They did not observe a peak in $\tilde{m}_2$ at the critical point of the 1D quantum Ising ferromagnet. 
Given that our results as well as those MPS/TN results are derived from direct calculations on the TFI model rather than the corresponding classical stochastic matrix form Hamiltonian (not an exact TFI model but has same phases), the peak behavior at the critical point of the 1D TFI model must be solid. 
On another hand, the results in Fig.~\ref{fig:gs_sre} reflect that the magnitude of magic is not a characteristic quantity for a phase or phase transition. For example, the state can be efficiently prepared using Clifford gates when the spins are strictly aligned along the $z$-axis. However, the magic increases if the spins are distributed around the north pole of the Bloch sphere, even though the state remains in the FM phase. Similarly, when the spins rotate to align precisely along the $x$-axis, characteristic of a perfect paramagnetic phase, the magic returns to zero. 
An important reason for the above observations is that magic is essentially basis-dependent, which can be viewed as the participation entropy in the Pauli basis~\cite{turkeshi2023paulispectrum,collura2024fermionmagic,frau2024magic_disentangling}. In this sense, its derivatives and scaling behavior could be more important. As we will discuss later, the volume-law corrections are also valuable.

\begin{figure}[ht!] \centering
    \includegraphics[width=8cm]{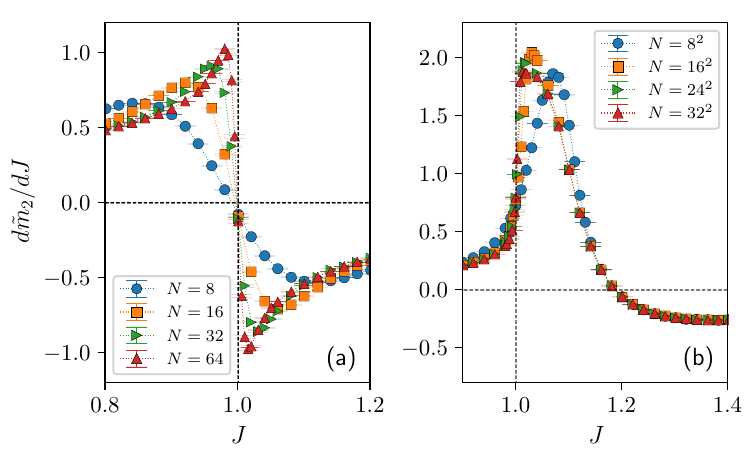}
    \caption{$d\tilde{m}_2/dJ$ of the TFI models: (a) 1D ring with length $L$ ;(b) 2D $L\times L$ square lattice.
    }
    \label{fig:gs_sre_dJ}
\end{figure}

\begin{figure}[ht!] \centering
    \includegraphics[width=8cm]{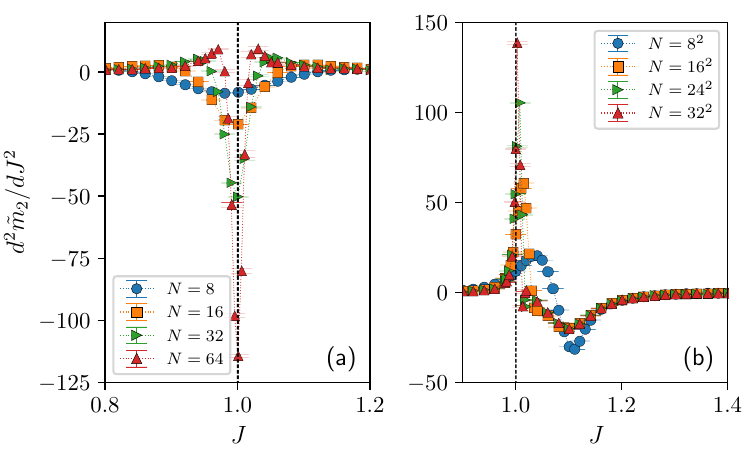}
    \caption{$d^2\tilde{m}_2/dJ^2$ of the TFI models: (a) 1D ring with length $L$ ;(b) 2D $L\times L$ square lattice.
    The lowest value is at $J\approx 1.1$.
    }
    \label{fig:gs_sre_d2J}
\end{figure}

Fig.~\ref{fig:gs_sre_dJ} and Fig.~\ref{fig:gs_sre_d2J} presents of the first- and second-order derivatives of $\tilde{m}_2$, respectively.
For both the 1D and 2D cases, singularities can be observed. Particularly, their second-order derivatives diverge at the critical points. From Eqs.(\ref{eq:first_order_J}) and (\ref{eq:second_order_J}), the singular behavior of the 2-SRE is determined by the interplay between the higher-order derivatives of the (logarithmic) partition functions $\log Q$ and $\log Z$. Notice that $Z_2$ and $Z$ have same behaviors of singularity at zero temperature. If the singularity originates purely from the $Z$-part (related to the free energy or energy), it is trivial, as both the 1D and 2D models exhibit second-order phase transitions at their quantum critical points. Consequently, if the $Q$-part, which is tied to the characteristic function $\Xi_P(\rho)$, exhibits no singularity, then magic lacks a direct sensibility to criticality for the model.

% ----------------------------------------------------
% ----------------------------------------------------
% ----------------------------------------------------

% =======================================
\subsection{Contributions from $Q$ and $Z$ for the singularities in the ground-state magic}\label{sec:qz_discuss}
\begin{figure}[ht!] \centering
    \includegraphics[width=8cm]{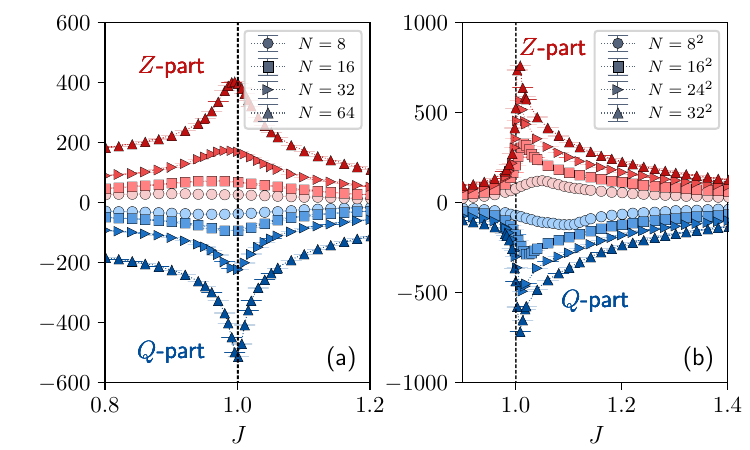}
    \caption{For $d^2\tilde{m}_2/dJ^2$, the contributions from $Q$ and $Z$ according to Eqs.(\ref{eq:first_order_J}) and (\ref{eq:second_order_J}): (a) the 1D ring with length $L$; (b) 2D $L\times L$ square lattice. 
    }
    \label{fig:gs_sre_qz}
\end{figure}
To clarify, we separately plot the $Q$- and $Z$-parts in Fig.~\ref{fig:gs_sre_qz}. For the TFI models, we surprisingly find that both parts exhibit divergence at $J_c$. In the 1D case, depicted in Fig.~\ref{fig:gs_sre_qz}(a), the $Q$-part governs the behavior of $[d^2\tilde{m}_2/dJ^2]$ near the critical point, resulting in the sharp downward trend of $[d^2\tilde{m}_2/dJ^2]$ in Fig.~\ref{fig:gs_sre_d2J}(a). By contrast, in the 2D case shown in Fig.~\ref{fig:gs_sre_qz}(b), the $Z$-part dominates. Away from the critical point, the divergent effects of $Q$- and $Z$-parts cancel out in the ferromagnet phase, leading to a distinct behavior of $[d^2\tilde{m}_2/dJ^2]$ in Fig.~\ref{fig:gs_sre_d2J}(b) with an extreme value at $J\approx 1.1$.

The results above suggest that in general models, the interplay between the $Q$- and $Z$-parts, or the contributions from the characteristic function and free energy, may result in more intricate behavior of $\tilde{m}_2$. The the order of the phase transition plays a significant role here. From this perspective, it is more clear that magic quantified by $\tilde{m}_2$ does not necessarily achieve a maximum or minimum at the quantum critical point consequently.
This argument is also supported by previous small-scale numerical results for the 2D $\mathbb{Z}_2$ gauge theory, obtained using tree tensor network sampling~\cite{poetri2023magic}. 
For the 1D and 2D TFI models studied here, the divergence of the $Q$-part in $[d^2\tilde{m}_2/dJ^2]$, which relates to $\Xi_P(\rho)$, suggests a more direct connection between magic and (conformal) criticality.
However, this relationship is still subtle, as the $Q$-part may lack a singularity for certain models and general bases.

% =======================================
\subsection{Volume-law corrections}\label{sec:volume}
\begin{figure}[ht!] \centering
    \includegraphics[width=8.5cm]{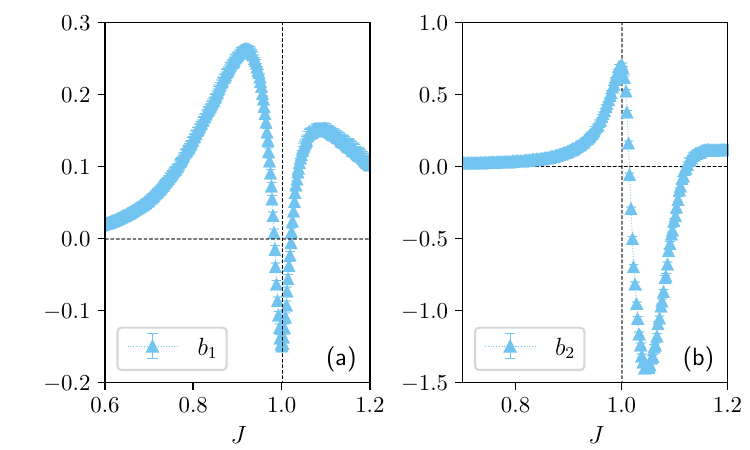}
    \caption{
        The volume law corrections for the ground states of (a) the 1D TFI model; (2) the 2D TFI model. The lowest point of $b_2(J)$ is at $J\approx 1.05$.
    }
    \label{fig:gs_sre_correction}
\end{figure}
To gain deeper insight into the results and many-body magic, we discuss the volume-law corrections of $2$-SRE. 
In the following discussion, we assume that the ground state exhibits translation symmetry, which holds in many cases including the TFI models considered in this work.
% For anisotropic quantum systems such as the Heisenberg spin glass model, however, the magical behavior of its ground state could be more complicated. 
If a symmetric ground state exhibits only local magic, for example, $\ket{\psi}=[(\ket{\uparrow\downarrow}-e^{i\pi/4}\ket{\downarrow\uparrow})/\sqrt{2}]^{\otimes N/2}$, which is a product state of $N/2$ singlet states up to a phase, the strict volume law $\tilde{M}_2\propto N$ of the full-state magic follows directly from the additivity property discussed in Sec.~\ref{subsec:se}. 
In contrast, if a ground state exhibits nonlocal magic, which cannot be removed via finite-depth local quantum circuits in the thermodynamic limit~\cite{Korbany2025longrange,Weifuchuan2025longrange}, then nonzero volume-law corrections must appear in the full-state magic.
These corrections have similar role with the mutual magic $\mathcal{L}_{AB}(\rho):=\mathcal{M}(\rho_{AB})-\mathcal{M}(\rho_{A})-\mathcal{M}(\rho_B)$~\cite{white2021cftmagic,Fliss2021long_range_magic,Sarkar_2020_1d_xy_rom,ning2022long_range_magic,frau2024magic_disentangling} in a bipartite system ${A}\cup{B}$, where $\mathcal{M}$ is some mixed-state magic monotone. Unlike mutual magic, these volume-law corrections go beyond the bipartition scenario.
Although nonzero volume-law corrections are a necessary but not sufficient condition for the presence of nonlocal magic, they serve as a valuable diagnostic tool, particularly given the difficulty of directly identifying nonlocal magic through circuit-based definitions.
This mirrors the difficulty of identifying long-range entanglement in many-body systems. 
For instance, in conformal critical systems, subleading corrections to the area law of entanglement entropy often serve as important indicators~\cite{Calabrese2009entcft,Nishioka2018entreview}. 
Similarly, we claim that, in the thermodynamic limit, persistent volume-law corrections to SRE at a conformal critical point may indicate nonlocal magic.

To fit the volume-law corrections, we adopt the fitting ansatz
\begin{equation}\label{eq:fittttt}
    \tilde{M}_2=a_dL^d+b_d
\end{equation}
for ground states on a $d$-dimensional lattice with length $L$ with the data of $\tilde{m}_2$ in Fig.~\ref{fig:gs_sre}. Additional subleading terms may be present, particularly in higher dimensions.
As shown in Fig.~~\ref{fig:gs_sre}, the volume-law coefficients $a_1$ and $a_2$ for the TFI models are close to their corresponding $2$-SRE density, confirming that volume law is exactly the leading term in $2$-SRE. 

Fig.~\ref{fig:gs_sre_correction}(a) presents the correction $b_1$ for the ground states of the 1D TFI model. Since both the paramagnetic phase and the ferromagnet phase are gapped, in the thermodynamic limit, no nonlocal magic should be captured by $2$-SRE. This reasoning also extends to the 2D case shown in Fig.~\ref{fig:gs_sre_correction}(b). Therefore, we attribute the nonzero values of $b_d$ ($d=1,2$) near $J_c$ to finite-size effects. More interestingly, these finite-size effects lead to the indication of discontinuities at the critical points. 
The difference between the two phases are also reflected by the discontinuities of the magical structures reflected by $b_d$.  
This phenomenon contrasts strongly with the full-state magic, from which extracting information of phases and criticalities proves challenging, as we discussed in Sec.~\ref{sec:qz_discuss}.
% Though the full-state magic is basis-dependent, the nonlocal correlations at the critical point could always require nonlocal non-Clifford operations. 
This suggest that the volume-law corrections of magic could provide a stronger diagnostic for criticalities compared to full-state magic. 
This is especially significant for $\alpha$-SRE, which, as a non-mixed-state measure, fails to provide a reliable or monotonic mutual magic for bipartite systems. 
% By contrast, studying volume-law corrections provide a practical means of quantifying nonlocal magic and criticality with $\alpha$-SRE. 

\begin{figure}[ht!] \centering
    \includegraphics[width=6cm]{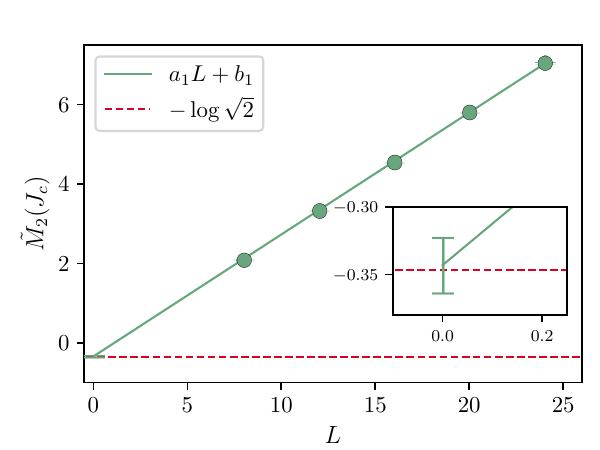}
    \caption{ 
        The volume-law scaling $\tilde{M}_2=a_1L+b_1$ fitted for the 2-SRE of the 1D TFI model (PBC) with length $L$ at its critical point $J_c=1$. The correction term is determined to be $b_1=-0.34(2)$, which is consistent with the theoretical prediction $-\log\sqrt{2}\approx -0.3466$ in Ref.~\cite{Hoshino2025bcft}.
        }
        \label{fig:correction1}
\end{figure}

Importantly, a recent theoretical study has shown that for a (1+1)D conformal critical point, the volume-law correction to the $2$-SRE is a constant~\cite{Hoshino2025bcft}.
This constant is connected to the non-integer ground-state degeneracy, known as the $g$-factor, of some boundary conformal field theory. 
In the 1D TFI model, this yields $b_{1,\mathrm{exact}}=-\log\sqrt{2}\approx -0.3466$. 
To verify this, we carried out low-temperature simulations with $\beta=4L$ and successfully extracted $b_1=-0.34(2)$, in excellent agreement with the theoretical expectation, as shown in Fig.~\ref{fig:correction1}.

% =======================================
\subsection{$2$-SRE of 2D finite-temperature Gibbs states}
% ---------------------------------------------------
\begin{figure}[ht!] \centering
    \includegraphics[width=8cm]{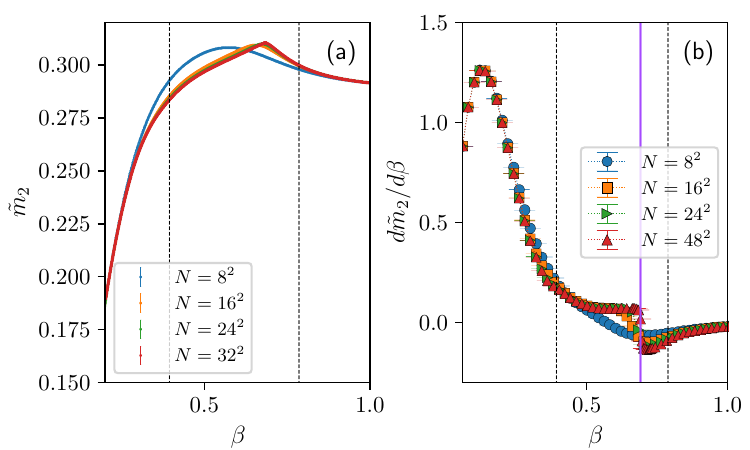}
    \caption{Fixing $J=1$ and $h=2.5$, for the 2D TFI model on a $L\times L$ square lattice: (a) variation of $\tilde{m}_2$ with the inverse temperature $\beta$; (b) the first-order derivative $d\tilde{m}_2/d\beta$. For each subfigure, the left dashed line refers to $\beta_c/2$ and the right one refers to $\beta_c$.
    The additional purple solid line in the right figure corresponds to the singular point where $\beta^*\approx 0.687$.
    }
    \label{fig:T_sre}
\end{figure}
While $\alpha$-SRE is not well-suited as a measure for mixed states, we are interested in examining whether it can partially reflect some of the system's properties.
We fix $J=1$ and $h=2.5$, then the finite-temperature phase transition point is located at $T_c=1.27369(5)$ or $\beta_c\approx 0.78512$ based on the QMC results in Ref.~\cite{Hesselmann2016TFIM_Tc}. 

At the infinite-temperature limit ($\beta= 0$), $\tilde{m}_2$ is approximately zero within the error bar. 
For finite temperatures, however, $\tilde{m}_2$  exhibits complex behavior, offering minimal valuable information, as shown in Fig.~\ref{fig:T_sre}(a). Additionally, for the first-order derivative $d\tilde{m}_2/d\beta$, we observe a singular point $\beta^*\approx 0.687<\beta_c$ in the paramagnetic phase, as shown in Fig.~\ref{fig:T_sre}(b).
% ---------------------------------------------------
\begin{figure}[ht!] \centering
    \includegraphics[width=8cm]{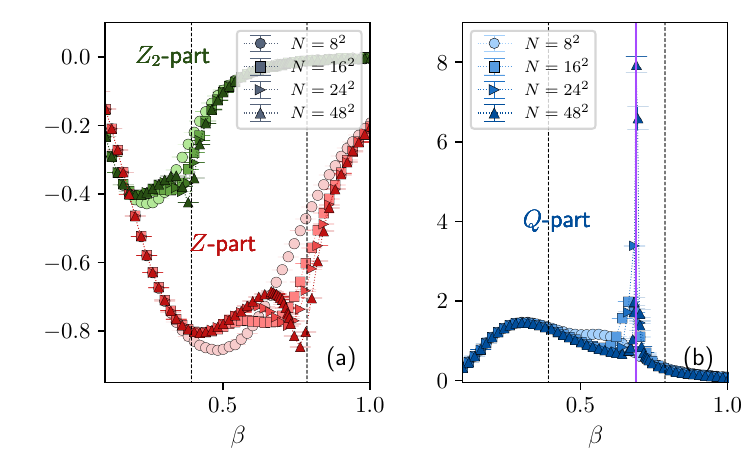}
    \caption{
        For $-\beta^2(d^2\tilde m_2/d\beta^2)$: (a) the $Z$ and $Z_2$ contributions (b) the $Q$ contribution.
    }
    \label{fig:T_sre_qzz2}
\end{figure}
% ---------------------------------------------------

Using Eq.~(\ref{eq:second_order_beta}), we analyze $-\beta^2(d^2\tilde m_2/d\beta^2)$, which is proportional to the specific heat. For this second-order phase transition, the $Z$-part of $-\beta^2(d^2\tilde m_2/d\beta^2)$ must be singular at $\beta_c$. Meanwhile, the $Z_2$-part, influenced by the doubling of $\beta$, attains its critical point at $\beta_c/2$, as shown in Fig.~\ref{fig:T_sre_qzz2}(a).
Unexpectedly, the $Q$-part, depicted in Fig.~\ref{fig:T_sre_qzz2}(b), diverges at $\beta^*$, a point unrelated to any physical property of the system.
Consequently, the $2$-SRE, which is not a well-defined magic measure for mixed states, proves inadequate for characterizing phases or critical phenomena at the finite-temperature case.

% ::::::::::::::::::::::::::::::::::::::::::::::::::::::::::
% ::::::::::::::::::::::::::::::::::::::::::::::::::::::::::
% ::::::::::::::::::::::::::::::::::::::::::::::::::::::::::
% ::::::::::::::::::::::::::::::::::::::::::::::::::::::::::
\section{conclusion and discussions}\label{sec:summary}
In this work, we introduce a novel and efficient QMC method for computing the $\alpha$-SRE and its derivatives in large-scale and high-dimensional quantum many-body systems. Our method is broadly applicable to any sign-problem-free Hamiltonian regardless of the dimension of the system. 
This versatility makes our method well suited for studying quantum magic across a wide range of physical systems, including those exhibiting spontaneous symmetry breaking, topological phase transitions, topological order, and conformal criticality. 
As many-body magic remains an underexplored frontier, our algorithm offers a vital computational tool for advancing this area, particularly in light of recent theoretical developments in conformal field theory~\cite{Hoshino2025bcft}.

We demonstrate our approach on both 1D and 2D ground states, as well as finite-temperature Gibbs states of the TFI model. 
A key benefit of our approach is that it allows us to isolate the free energy contribution in $\alpha$-SRE and the derivatives, providing deeper understanding into magic and criticality in quantum many-body systems, which are inaccessible before. 
Particularly, we uncover singularities in the derivatives of the characteristic function contribution in SRE at the conformal critical points in both 1D and 2D.
Further, we reveal that the behavior of the $\alpha$-SRE is governed by the interplay between these two contributions, clarifying why full-state magic generally fails to reveal meaningful information about phase structure and criticality, as observed in earlier studies.
Meanwhile, the order of the phase transition plays a crucial role for the behaviors of SRE and the singularities in its derivatives.
This is quite different from quantum entanglement: for both the 1D and 2D TFI model, previous results show that the quantum entanglement reaches its maximum at the quantum critical points~\cite{Calabrese2004ent_qft,wang2024ent_entropy}. 

To further investigate the use of full-state magic, we consider the volume-law corrections, whose nonzero values serve as a necessary condition for the presence of nonlocal magic inherent in system correlations. 
These corrections, observed in our simulations on finite system sizes, show the evidence of discontinuity at quantum critical points, indicating distinct magical structures on the two sides of the critical point. 
We consider volume-law corrections to be more significant, and argue that they could be important diagnostics for nonlocal magic at conformal critical points.
We leave a more comprehensive exploration of these corrections across a broader class of systems to future work. 
Finally, we confirm that $\alpha$-SRE is not an effective measure for mixed-state magic in many-body systems by studying an example of finite-temperature phase transition.

In addition, our algorithm shows not only certain magic states of a high-dimensional many-body Hamiltonian can be simulated efficiently on a classical computer, but their magic can be computed efficiently. This underscores the fact that magic is just a necessary but not sufficient resource for quantum advantage.

In closing, we discuss extending our algorithm to compute the magic of a reduced density matrix.
The mutual magic $\mathcal{L}_{AB}$ quantifies those magic generates from nonlocal non-Clifford operations, and it could also encode crucial information of many-body systems. Previous studies have shown that the mutual magic $\mathcal{L}_{AB}$ would exhibit stronger signatures than the full-state magic in detecting criticalities and is capable of characterizing certain states in topological quantum field theory~\cite{white2021cftmagic,Fliss2021long_range_magic,Sarkar_2020_1d_xy_rom, poetri2023magic, montana2024long_range_sre,ning2022long_range_magic, frau2024magic_disentangling}. 
While $\mathcal{L}_{AB}$ is ideally defined using a mixed-state magic monotone, using $\alpha$-SRE to define it, called the mutual stabilizer R\'enyi entropy (mSRE), has also proven valuable~\cite{poetri2023magic,frau2024magic_disentangling,montana2024long_range_sre}, much like quantum mutual information captures entanglement. 
Its significance has also been theoretically substantiated in the recent development on conformal field theory~\cite{Hoshino2025bcft}.
Our algorithm can be easily extended to compute reduced density matrix magic by tracing out one subsystem in QMC simulations using the technique in~\cite{Yan2023wormhole,lch2024ent_specturm,mbb2023sampling,Wang2025sampling}, so that the mSRE can be computed. This is also a promising way to study the finite-temperature phase transition and open quantum systems rather than considering the full-state $\alpha$-SRE, as mentioned at the beginning. Our work will enhance the further fusion and intersection of quantum information and many-body physics.

% ::::::::::::::::::::::::::::::::::::::::::::::::::::::::::
% ::::::::::::::::::::::::::::::::::::::::::::::::::::::::::
% ::::::::::::::::::::::::::::::::::::::::::::::::::::::::::
% ::::::::::::::::::::::::::::::::::::::::::::::::::::::::::
\section*{Acknowledgements}
We thank P.S. Tarabunga, M. Dalmonte, E. Tirrito, H. Timsina, T. Byrnes, C. Radhakrishnan, K. Warmuz, Z. Liu, G.-Y. Zhu, Y. Tang, and Z.-Y. Wang for helpful discussions. Z.W. acknowledges the China Postdoctoral Science Foundation under Grant No.~2024M752898.  The work is supported by the Scientific Research Project (No.WU2024B027) and the Start-up Funding of Westlake University. The authors also acknowledge the HPC centre of Westlake University and Beijng PARATERA Tech Co.,Ltd. for providing HPC resources.start-up funding of Westlake University. 

% ---------------------------------------------------------
% ---------------------------------------------------------
% ---------------------------------------------------------
% ---------------------------------------------------------
\appendix
% ---------------------------------------------------------
\section{Nonlocal update and autocorrelation time for simulating $Q$ in the TFI model}\label{appx:tfim}
\begin{figure}[ht!] \centering
    \includegraphics[width=10cm]{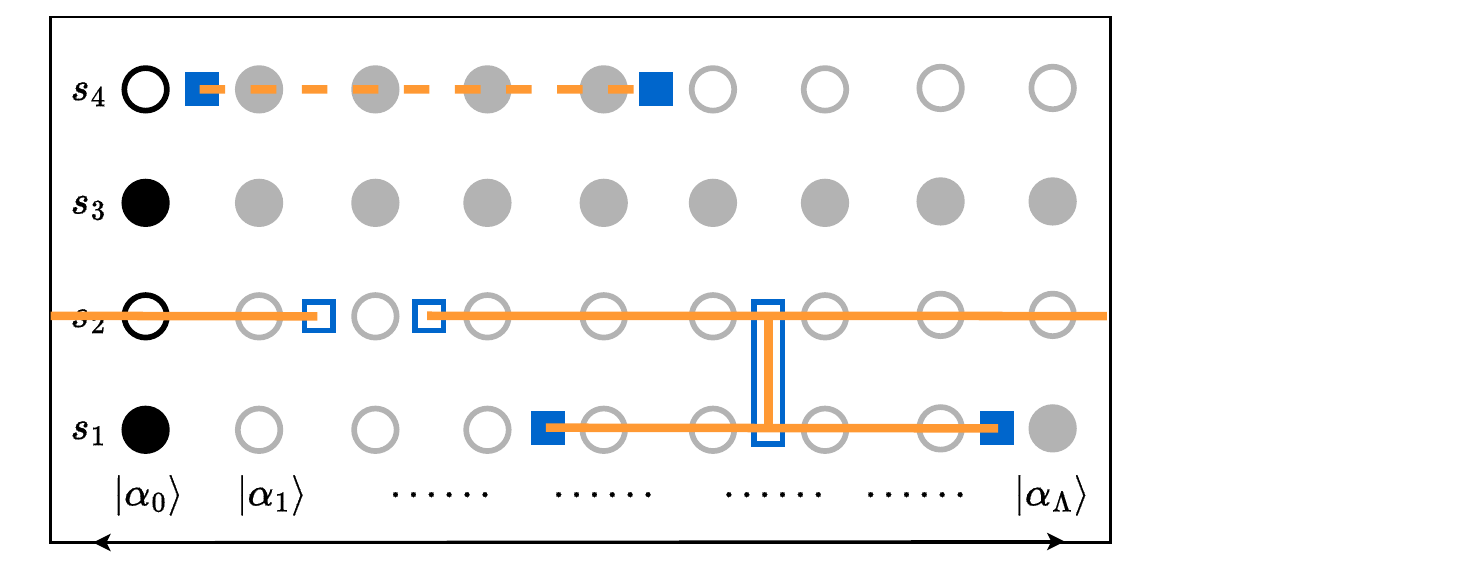}
    \caption{ 
        To grow a cluster, we begin from either a site or a bond operator and identify all operators belonging to the cluster based on the following rules: (1) each cluster terminates at site operators, $H_{-1,j}$ or $H_{0,j}$; (2) each bond operator $H_{1,\overline{jk}}$ is associated with a single cluster. The example illustrates two clusters, represented by dashed and solid cyan lines, respectively.
        }
        \label{fig:branch_cluster}
\end{figure}

For a conventional SSE simulation of partition function $Z$ of TFI model, we define the following operators~\cite{sandvik2003stochastic,Melko2013}
\begin{align}
    &H_{-1,j}=h(\sigma_j^++\sigma_j^-) , \\ 
    &H_{0, j} = h I_j, \\
    &H_{1, \overline{jk}} = J(Z_jZ_k+I_jI_k) ,
\end{align}
where $\sigma^{\pm}:=(X\pm iY)/2$. 
The nonlocal update in this case is the branching cluster update, as illustrated in Fig.~\ref{fig:branch_cluster}, where each cluster is updated with a $50\%$ probability by switching between single-site diagonal operators $H_{0,j}$ and off-diagonal operators $H_{-1,j}$. If a cluster crosses the imaginary-time boundary, as the cluster represented by the solid orange lines in Fig.~\ref{fig:branch_cluster}, we should accordingly update the spins in $\ket{\alpha_0}$.

\begin{figure}[ht!] \centering
    \includegraphics[width=9cm]{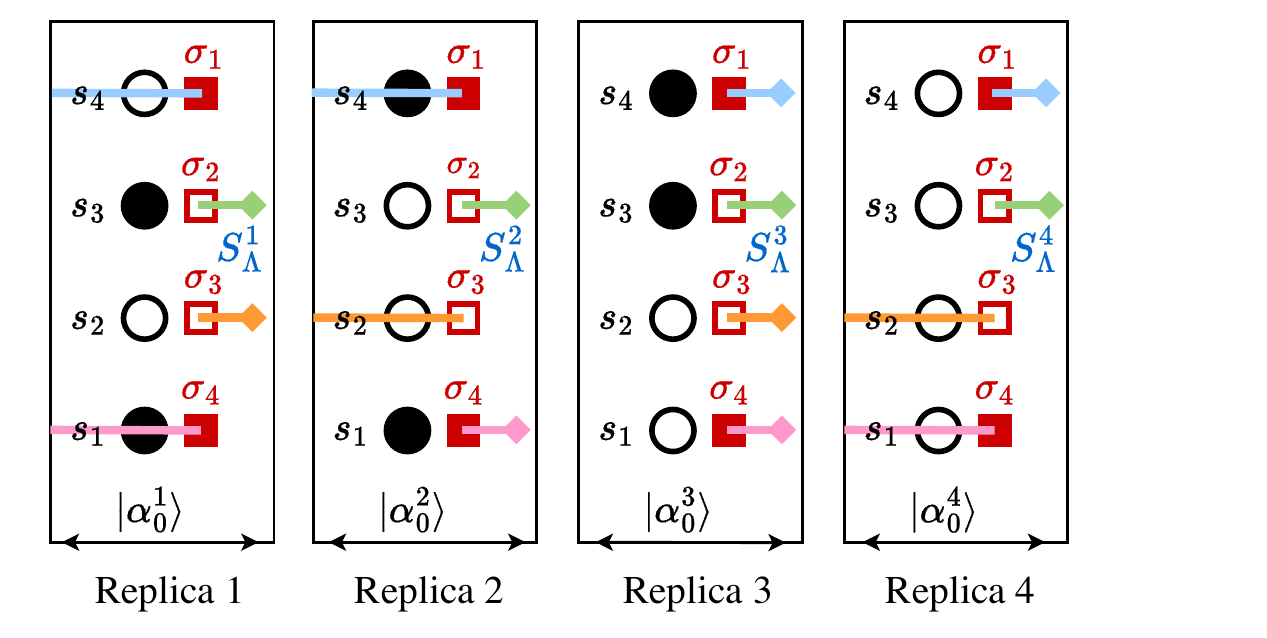}
    \caption{ 
        For each site $s_j$, since $\sigma_j$ in different replicas must be the same, a cluster associated with $s_j$ must include all the $\sigma_j$. For each $s_j$, we start at either the left or right side of $\sigma_j$ in each replica, but ensure that the imaginary-time boundaries are crossed an even number of times in total. The diamond-shaped markers represent the remaining part of the cluster, which is composed of $H_{-1,j}$, $H_{0,j}$ and $H_{1,\overline{jk}}$. 
        Notably, due to the existence of the two-site operators $H_{1,\overline{jk}}$, different $\sigma_j$ and $\sigma_k$ can belong to the same cluster.
        }
    \label{fig:cross_replica_branch_cluster}
\end{figure}

As introduced in Sec.~\ref{subsec:algorithm}, to simulate $Q$, we extend the clusters defined in Fig.~\ref{fig:branch_cluster} to additionally include $\{\tilde{\sigma}_j\}$ from the reduced Pauli string $\tilde{P}=\prod_j\tilde{\sigma}_j$, and allow the cluster to span multiple replicas (see Fig.~\ref{fig:cross_replica_branch_cluster}). 
While the update scheme presented above appears reasonable, we find that, in practice, it results in long autocorrelation times. This occurs because the cross-replica clusters inadvertently correlate the operators in different replicas, which are supposed to be independent.
To fix this problem, before each round of nonlocal updates, we freeze each $\sigma_j$ with a probability of $50\%$ to make the operators in different clusters independent enough.
Apparently, the detailed balance condition still holds with this modification. If a $\sigma_j$ is frozen, then it will not be included in any cluster. Unlike the clusters in Fig.~\ref{fig:cross_replica_branch_cluster} which consider four replicas, only two replicas are randomly chosen in this case related to the site $j$. Additionally, the imaginary-time boundaries in the selected two replicas are either crossed simultaneously or not at all. We find these modifications lead to a dramatic reduction in autocorrelation times to an acceptable level.

\begin{figure}[ht!] \centering
    \includegraphics[width=8cm]{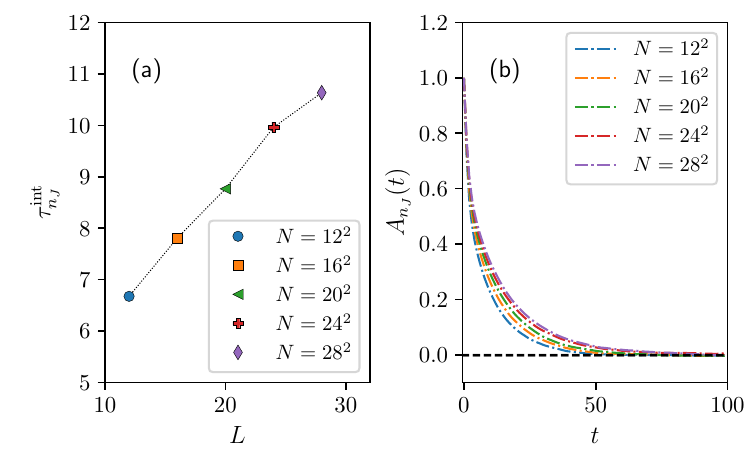}
    \caption{ 
       (a) The integrated autocorrelation time of $n_J$ for the 2D TFI model at the quantum critical point with $h/J=3.0443$ and $\beta=L$ is shown; (b) The decay of $A_{n_J}(t)$ for various system sizes, represented by different colors.
    }\label{fig:autocorrelation}
\end{figure}

As an example, we consider the integrated autocorrelation time of the observable $n_J$ in Sec.~\ref{eq:re_an_intro}, defined as
\begin{equation}\label{eq:autotime}
    \tau_{n_J}^{\text{int}}:=\frac{1}{2}+\sum_{t=1}^{\infty} A_{n_J}(t) ,
\end{equation}
where 
\begin{equation}\label{eq:autofunc}
    A_{n_J}(t):=\frac{\langle n_J(u)n_J(u+t)\rangle-\langle n_J(u)\rangle^2}{\langle n_J(u)^2\rangle-\langle n_J(u)\rangle^2}
\end{equation}
is the autocorrelation function with $u$ and $t$ to denote different MC times~\cite{evertz2003loopAlgorithm, sandviki2011tutorial}.
$A_{n_J}(t)$ is a convex function and will decay exponentially when $t$ is large, and the sum in Eq.~(\ref{eq:autotime}) is truncated when $A_{n_J}(t)$ becomes negligibly small in practice~\cite{evertz2003loopAlgorithm}. 
The results are shown in Fig.~\ref{fig:autocorrelation}, which indicate the efficiency of this modified algorithm for nonlocal updates.

% ---------------------------------------------------------
\section{Estimators in reweight-annealing}\label{appx:estimator}
We consider the partition function $Q$ as an example, with similar reasoning applying to $Z$ and $Z_2$. Suppose $Q(\lambda_k)=\sum_c W_c(\lambda_k)$, where $W_c(\lambda_k)\ge 0$ are the real weights, then 
\begin{equation}\label{eq:reweighting_trick}
    \begin{split}
        \frac{Q(\lambda_{k-1})}{Q(\lambda_k)} 
    =&\frac{\sum_c W_c(\lambda_{k-1})}{\sum_c W_c(\lambda_{k})}  \\ 
    = & \frac{\sum_c \frac{W_c(\lambda_{k-1})}{W_c(\lambda_k)}W_c(\lambda_{k})}{\sum_c W_c(\lambda_{k})} \\ 
    =&\bigg\langle  \frac{W(\lambda_{k-1})}{W(\lambda_k)} \bigg\rangle_{\lambda_k,Q} ,
    \end{split}
\end{equation}
which averages the weight ratios for the two parameters $\lambda_{k-1}$ and $\lambda_k$. 
Eq.~(\ref{eq:reweighting_trick}) is called the reweighting trick~\cite{Ferrenberg1988reweighting,ding2024reweight}. 
If $\lambda\equiv\beta$, then according to Eq.~(\ref{eq:Q_weight}) and Eq.~(\ref{eq:prop_rep}), the ratio $W(\beta_{k-1})/W(\beta_k)$ in Eq.~(\ref{eq:reweighting_trick}) becomes $(\beta_{k-1}/\beta_k)^{n_{\text{tot}}}$, where $n_{\text{tot}}$ is the total number of operators in the four replicas.
Similarly, we can obtain the reweighting ratio when $\lambda\equiv J$.

\bibliography{ref}

\end{document}